\documentclass[entropy,article,submit,moreauthors,10pt,a4paper]{mdpi}
\firstpage{1}
\articlenumber{x}
\doinum{10.3390/------}
\pubvolume{xx}
\pubyear{2017}
\copyrightyear{2017}
\pdfoutput=1

\usepackage{amssymb}
\usepackage{amsfonts}

\makeatletter
\newcommand{\subalign}[1]{%
  \vcenter{%
    \Let@ \restore@math@cr \default@tag
    \baselineskip\fontdimen10 \scriptfont\tw@
    \advance\baselineskip\fontdimen12 \scriptfont\tw@
    \lineskip\thr@@\fontdimen8 \scriptfont\thr@@
    \lineskiplimit\lineskip
    \ialign{\hfil$\m@th\scriptstyle##$&$\m@th\scriptstyle{}##$\crcr
      #1\crcr
    }%
  }
}
\makeatother

\newcommand\custdots{\makebox[1em][c]{.\hfil.\hfil.}\,,\makebox[1em][c]{.\hfil.\hfil.}}

\DeclareMathOperator{\sech}{sech}
\DeclareMathOperator{\arsinh}{arsinh}
\DeclareMathOperator{\erf}{erf}
\DeclareMathOperator{\erfi}{erfi}

 \theoremstyle{mdpi}
 \newcounter{thm}
 \newcounter{ex}
 \newcounter{re}
 \theoremstyle{mdpidefinition}

\Title{Taxis of Artificial Swimmers in a Spatio-Temporally Modulated Activation Medium}

\Author{Alexander Geiseler $^{1,*}$, Peter H\"anggi $^{1,2,3}$, and Fabio Marchesoni $^{4,5}$}
\AuthorNames{Alexander Geiseler, Peter H\"anggi, and Fabio Marchesoni}

\address{%
$^{1}$ \quad Institut f\"ur Physik, University of Augsburg, D-86159, Germany\\
$^{2}$ \quad Nanosystems Initiative Munich, Schellingstraße 4, D-80799 M\"unchen, Germany\\
$^{3}$ \quad Department of Physics, National University of Singapore, 117551 Singapore, Republic of Singapore\\
$^{4}$ \quad Center for Phononics and Thermal Energy Science, School of Physics Science and Engineering, Tongji University,
Shanghai 200092, People's Republic of China\\
$^{5}$ \quad Dipartimento di Fisica, Universit\`{a} di Camerino, I-62032 Camerino, Italy\\}

\corres{Correspondence: alexander.geiseler@physik.uni-augsburg.de}

\abstract{Contrary to microbial taxis, where a tactic response to external stimuli is controlled by complex chemical pathways acting like sensor-actuator loops, taxis of artificial microswimmers is a purely stochastic effect associated with a non-uniform activation of the particles' self-propulsion. We study the tactic response of such swimmers in a spatio-temporally modulated activating medium by means of both numerical and analytical techniques. In the opposite limits of very fast and very slow rotational particle dynamics, we obtain analytic approximations that closely reproduce the numerical description. A swimmer drifts on average either parallel or anti-parallel to the propagation direction of the activating pulses, depending on their speed and width. The drift in line with the pulses is solely determined by the finite persistence length of the active Brownian motion performed by the swimmer, whereas the drift in the opposite direction results from the combination of ballistic and diffusive properties of the swimmer's dynamics.}


\conferencetitle{XXV Sitges Conference on Statistical Mechanics}

\begin{document}

\section{Introduction}
\label{Intro}

The directed movement of microorganisms, such as bacteria or cells, induced by an external stimulus is called taxis. It is categorized based on the nature of the stimulus and on whether the microorganisms head toward (positive taxis) or away (negative taxis) from the stimulus' source \cite{Murray1993}. Commonly, taxis is induced by certain chemicals (\emph{chemotaxis}) or light (\emph{phototaxis}), but alternative tactic mechanisms are also known, like \emph{rheotaxis}, the response to fluid flows, or \emph{gravitaxis}, the response to the gravitational field \cite{Armitage1999}. Taxis plays a major role in many biological processes, e.g., in the formation of cell layers and other biological structures. Moreover, many bacteria profit from pronounced tactic capabilities in their search for food or escape from toxic substances \cite{Adler1966, Berg2004}. They do so by means of a built-in chemical signaling network, which elaborates their physiological response to external stimulus gradients \cite{Wadhams2004}.

A biomimetic counterpart of microbial motility is the self-propulsion of artificial microswimmers, synthetically fabricated microparticles that propel themselves by converting an external activating ``fuel'' into kinetic energy \cite{Schweitzer2003, Walther2013, Elgeti2015Rev, Bechinger2016}. Under certain operating conditions, such particles generate local non-equilibrium conditions in the suspension medium, which in turn exerts on them a thermo- \cite{Wurger2007, Jiang2010, Buttinoni2012, Yang2013}, electro- \cite{Moran2010, Ebbens2014}, or diffusiophoretic \cite{Golestanian2005, Howse2007, Volpe2011} push. Because the ability to control the transport of such particles is emerging as a key task in nanorobotic applications, rectification of artificial microswimmers is currently the focus of intense cross-disciplinary research. Unlike biological microorganisms, simple artificial mircoswimmers lack any internal sensing mechanism and thus cannot detect an activation gradient, with their response to the activating stimulus being instantaneous. Nevertheless, over the past few years, artificial microswimmers have been reported to undergo a tactic drift when exposed to static stimuli \cite{Hong2007, Ghosh2015, TenHagen2014, Uspal2015a, Lozano2016}.
In biological systems, however, tactic stimuli are seldom static, but more frequently modulated in the form of spatio-temporal signals, like traveling wave pulses. Some microorganisms are capable of locating the pulse source and heading toward it \cite{Armitage1990, Wessels1992}. This is an apparently paradoxical effect, because one expects rectification to naturally occur in the opposite direction, irrespective of the microorganisms' tactic response to a monotonic gradient. Indeed, assuming a symmetric pulse waveform, a microorganism orients itself parallel to the direction of the pulse propagation on one side of the pulse, and opposite to it on the other side. As the swimmer spends a longer time within the pulse when moving parallel to it, one would then expect it to "surf" the pulse and effectively move away from the pulse's source (Stokes' drift \cite{Stokes1847, Broeck2007}).
Experimental evidence to the contrary has been explained by invoking a finite adaption time of the mircoorganisms' response to temporally varying stimuli \cite{Hofer1994, Goldstein1996}.

By analogy with the taxis of ``smart'' adaptive biological swimmers, in a recent paper \cite{Geiseler2016PRE} we investigated the question of whether similar effects can be observed also for ``dumb'' artificial swimmers, that is, we considered a self-propelled particle subjected to traveling activation wave pulses. We numerically found that the particle drifts on average either parallel or anti-parallel to the incoming wave, the actual direction depending on the speed and width of the pulses. This behavior is a consequence of the spatio-temporal modulation of the particle's self-propulsion speed within the activating pulses. We complement now that first report by deriving new analytical results for the tactic drift of an artificial swimmer. For this purpose, in Sec.\ \ref{Summary} we review the results of Ref.\ \cite{Geiseler2016PRE}. In Sec.\ \ref{Results} we then focus on two limiting cases of the swimmer's dynamics, where an analytical treatment is viable. We conclude with a brief r\'{e}sum\'{e} in Sec.\ \ref{resume}.

\section{Artificial Microswimmers Activated by Traveling Wave Pulses}
\label{Summary}

At low Reynolds numbers, the dynamics of an artificial microswimmer diffusing on a 2D substrate and subjected to a spatio-temporally modulated activation can be modeled by the Langevin equations (LE) \cite{Geiseler2016PRE}

\begin{alignat}{3}
\label{LE}
&\dot{x}&&=v(x,t)\cos\phi&&+\sqrt{D_0}\,\xi_x(t),\nonumber\\
&\dot{y}&&=v(x,t)\sin\phi&&+\sqrt{D_0}\,\xi_y(t),\\
&\dot{\phi}&&=\sqrt{D_\phi}\,\xi_\phi(t).\nonumber
\end{alignat}
Here, $v(x,t)$ is the particle's self-propulsion velocity and $\phi$ denotes its orientation measured with respect to the $x$ axis. The above dynamics comprises three additive fluctuational noise sources---two translational of intensity $D_0$ and one rotational of intensity $D_\phi$---which, for simplicity, are represented by white Gaussian noise processes with zero mean and autocorrelation functions $\langle\xi_i(t)\xi_j(0)\rangle=2\delta_{ij}\delta(t)$ for $i,j=x,y,\phi$, as usually assumed in the current literature \cite{Bechinger2016}. The noises $\xi_i(t)$ model the combination of independent fluctuations, namely the thermal fluctuations in the swimmer's suspension fluid and the fluctuations intrinsic to its self-propulsion mechanism. Therefore, in the following we treat $D_0$ and $D_\phi$ as independent parameters. We remind that in the presence of the sole thermal fluctuations, for a spherical particle of radius $R$ the translational and rotational diffusion constants are related, that is, $D_0/D_\phi= 4R^2/3$ \cite{Serdyuk2007}.

When the swimmer's activation is not modulated, its self-propulsive velocity is nearly constant, i.e., $v(x,t)\to v_0$, and the particle performs an active Brownian motion with persistence time $\tau_\phi=D_\phi^{-1}$ and corresponding persistence length $l_\phi=v_0\tau_\phi$. On short timescales, its dynamics is then characterized by a directed \emph{ballistic} motion and on long timescales by an enhanced diffusion with zero shift and diffusion constant $\lim_{t\to\infty}\langle[x(t)-x(0)]^2\rangle/(2t)=D_0+D_s$, where $D_s=v_0^2/(2D_\phi)$ \cite{tenHagen2011}.

\begin{figure}[htbp]
\centering
\includegraphics[width=0.9\textwidth]{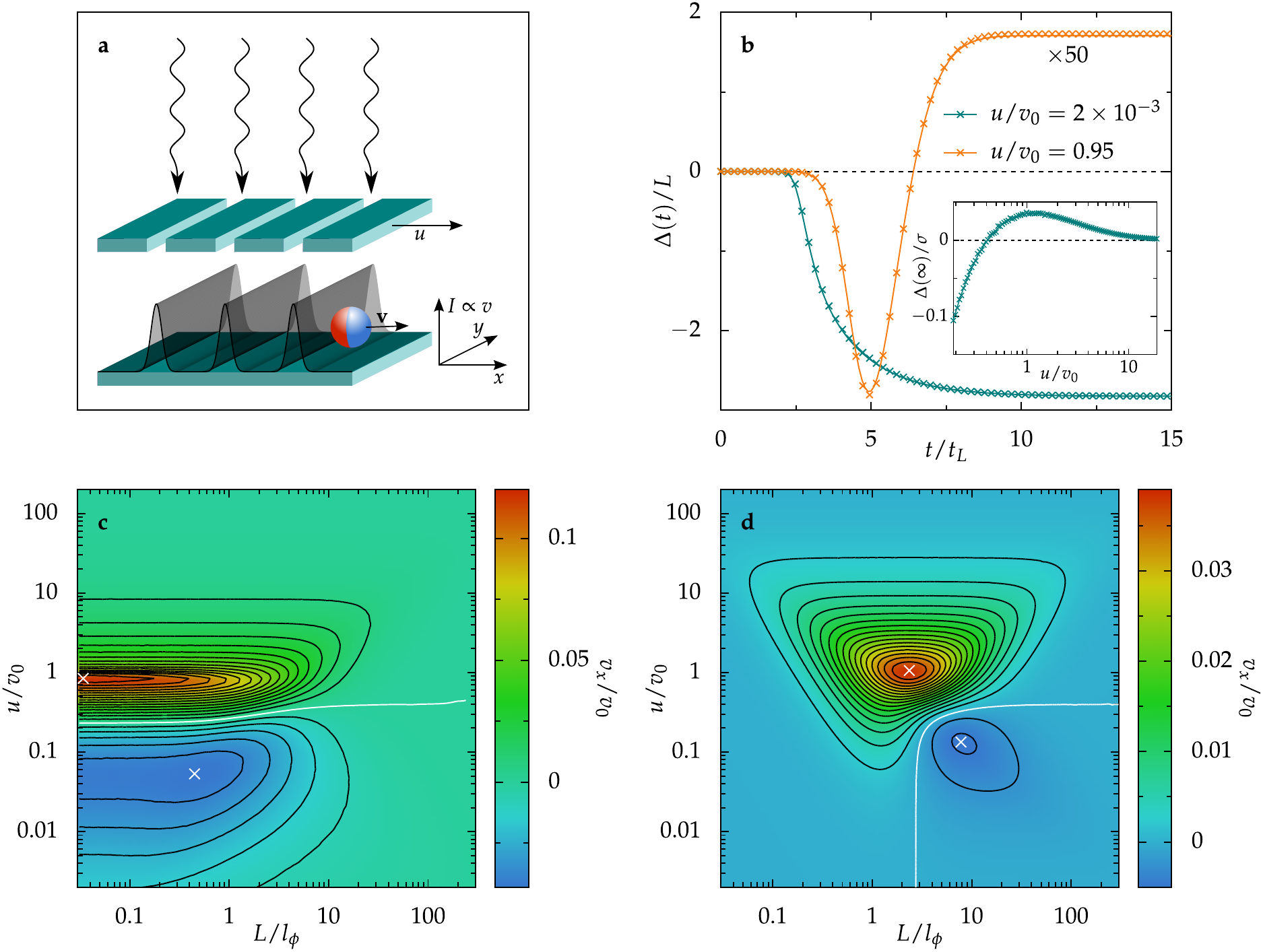}
\caption{Taxis of an artificial microswimmer subjected to traveling activation pulses. (\textbf{a}) Model setup to experimentally realize activating wave pulses as considered in the present paper, see text. (\textbf{b}) Tactic shift $\Delta(t)$ of the swimmer's mean position generated by a Gaussian activation pulse, $v(x,t)=v_0\exp[-(x-ut)^2/(2L^2)]$, vs.\ time $t$ in units of the pulse crossing time $t_L=L/u$. In the inset, the final shift $\Delta(\infty)$ is plotted as a function of the pulse speed $u$. The swimmer's self-propulsion parameters were set to $v_0=53\,\mu\mathrm{m}/\mathrm{s}$ and $D_\phi=165\,\mathrm{s}^{-1}$, and the pulse width was chosen according to $L=1\,\mu\mathrm{m}$, about three times the swimmer's propulsion length $l_\phi=v_0/D_\phi$. Here, the translational noise intensity $D_0$ was set to zero in order to focus on the essential mechanism giving rise to the swimmer's tactic shift. (\textbf{c},\textbf{d}) Tactic drift $v_x$ induced by a sinusoidal activation pulse, $v(x,t)=v_0\sin^2[(x-ut)\pi/L]$. The swimmer's parameters are the same as in (\textbf{b}) and we set $D_0=0$ in (\textbf{c}) and $D_0=2.2\,\mu\mathrm{m}^2/\mathrm{s}$ in (\textbf{d}). The position of the maximum positive drift and the maximum negative drift, respectively, are marked by white crosses and the white contours depict the separatrices dividing the regions of positive and negative taxis. All results were obtained either by stochastic integration of the LEs (\ref{LE}) [(\textbf{b}), crosses and (\textbf{c})] or by solving the corresponding FPE (\ref{FPEscaled}) [(\textbf{b}), solid lines and (\textbf{d})], see Ref.\ \cite{Geiseler2016PRE} for numerical details.\label{F1}}
\end{figure}

In Eq.\ (\ref{LE}) we assumed the swimmer's self-propulsion velocity, $v(x,t)$, to be a local function of the activating ``fuel'' concentration, which in turn can be modulated in time and space. An ideal setup allowing for the creation of traveling activation pulses is illustrated in Fig.\ \ref{F1}(\textbf{a}). In this sketch a thermophoretic swimmer activated by laser light \cite{Jiang2010, Volpe2011} is placed on a 2D substrate. Traveling wave pulses of laser intensity $I$ can be generated by sliding at constant speed $u$ a slit screen placed between the laser source and the particle. Because in a wide range of $I$ the swimmer's self-propulsive velocity is approximately proportional to the laser intensity \cite{Buttinoni2012}, one thus can generate any desired profile for $v(x,t)$. Although this is probably the simplest way to experimentally realize traveling activation pulses, we remark that chemically activated swimmers represent a viable option, too. Indeed, such swimmers can be operated under the condition that $v(x,t)$ is proportional to the concentration of the activating chemical(s), whereas their rotational diffusivity remains almost constant \cite{Howse2007, Hong2010}. On the other hand, traveling chemical waves can be conveniently excited in chemical reactors \cite{Kapral1995, Thakur2011, Lober2014}.

The effect of a single Gaussian activation pulse, $v(x,t)=v_0\exp[-(x-ut)^2/(2L^2)]$, hitting the swimmer from the left is depicted in Fig.\ \ref{F1}(\textbf{b}). Clearly, a pulse speed $u\ll v_0$ causes the particle to shift to the left, $\Delta(t):=\langle x(t)-x(0)\rangle<0$, whereas a pulse speed of about the same magnitude as the swimmer's maximum propulsion speed, $v_0$, causes it to shift slightly to the right. Indeed, we observe the final shift in the particle's position, $\Delta(\infty)=\lim_{t\to\infty}\Delta(t)$, to attain a positive maximum at $u\simeq v_0$ and tend toward large negative values for $u\to0$. As discussed in more detail in Sec.\ \ref{Results}, $\Delta(\infty)$ actually diverges in this limit if translational noise is neglected, $D_0=0$.

The existence of two opposing tactic regimes can be explained by considering the modulation of the swimmer's dynamics under the wave crests. Assuming no translational fluctuations, $D_0=0$, the swimmer can only diffuse within the pulse and comes to rest outside of it. For slow pulses, $u\ll v_0$, it propels very fast (compared to the pulse speed) in the wave center and thus quickly hits either pulse's edges, defined as the points where $u$ equals $v(x,t)$. Due to its movement to the right, the pulse's symmetry is dynamically broken and the two edges are not equivalent: if the swimmer crosses the right edge, it becomes slower than $u$ and is recaptured by the traveling pulse, whereas, by the same argument, it is left behind by the pulse once it crosses the left edge. The right (left) edge thus behaves like a reflecting (absorbing) boundary, which allows the particle to exit the pulse on the left only, hence inducing a negative tactic shift.
For pulse speeds approaching $v_0$, a contrasting effect comes into play: within the pulse, the particle can travel a longer distance to the right than to the left. This ``surfing'' behavior, already mentioned in Sec.\ \ref{Intro}, is most pronounced at $u=v_0$, where the distance a swimmer can travel to the right without hitting a pulse edge is solely limited by its rotational diffusivity, $D_\phi$. Accordingly, $\Delta(\infty)$ turns positive if $u$ becomes comparable to $v_0$ and vanishes monotonically in the limit $u\to\infty$, where the pulse sweeps through the swimmer so fast that it cannot respond. We note that the latter argument holds also for $D_0\neq 0$; as discussed in Sec.\ \ref{Results}, translational noise tends to suppress the swimmer's tactic shift, though not completely.

In Figs.\ \ref{F1}(\textbf{c},\textbf{d}), we consider a periodic sequence of pulses, namely $v(x,t)=v_0\sin^2[(x-ut)\pi/L]$, and measure the resulting steady-state tactic drift $v_x=\lim_{t\to\infty}\langle\dot{x}\rangle$ of the swimmer. Again, we keep the particle parameters $v_0$ and $D_\phi$ fixed and vary the wave parameters $L$ and $u$. In the absence of translational noise, Fig.\ \ref{F1}(\textbf{c}), we see essentially the same effect as in the case of a single activation pulse: $v_x$ is negative for $u\ll v_0$ and turns positive as $u$ approaches $v_0$, exhibiting a pronounced maximum at $u\simeq v_0$. However, the ratio between the maximum strength of the positive and negative tactic velocity, respectively, appears to be inverted. (For a single pulse, the negative shift at low $u$ is markedly larger than the positive shift at $u\simeq v_0$.) To this regard, we remind that in Fig.\ \ref{F1}(\textbf{c}) we plotted the net tactic drift, i.e., the speed defined as an average tactic shift divided by the relevant observation time. Since the large negative shift in Fig.\ \ref{F1}(\textbf{b}) occurs over a long time (the time needed by the swimmer to fully cross the Gaussian pulse is proportional to $L/u$), we expect the tactic drift velocity in Fig.\ \ref{F1}(\textbf{c}) to be less pronounced in the negative regime.
Moreover, we note that for pulse wavelengths $L$ larger than the swimmer's persistence length $l_\phi$, the action of the rotational noise becomes appreciable, leading to a suppression of $v_x$. This behavior is clearly consistent with Eq.\ (\ref{FPEscaled}), where for $D_0=0$ an increase in $L$ is equivalent to an increase in $D_\phi$.

As illustrated in Fig.\ \ref{F1}(\textbf{d}), translational fluctuations, $D_0>0$, suppress the tactic drift of the swimmer as well, because they help it diffuse across the wave troughs in both directions. Also the particle's ``surfing'' effect becomes less efficient and the tactic speed, $v_x$, diminishes overall. However, we notice that the translational noise has a stronger impact for small values of $L$, where it drastically suppresses the negative drift. This causes a sharp down-bending of the separatrix curve that divides the regions of positive and negative taxis, in correspondence with a critical value of $D_0/(Lv_0)$ \cite{Geiseler2016PRE}. As a matter of fact, one sees immediately that a decrease in $L$ is equivalent to an increase in $D_0$, since it is easier for the translational noise to kick a swimmer out of a pulse of smaller width. By the same argument it is also evident that translational fluctuations impact negative taxis more strongly than positive taxis. Indeed, the mechanism responsible for the negative drift requires preventing the swimmer from crossing a wave trough from left to right, which grows less efficient with increasing $D_0$.

We furthermore stress that in Eqs.\ (\ref{LE}) we neglected hydrodynamic effects, which, at least in the absence of activation gradients, are strongly suppressed by (i) restricting the swimmers' motion to the bulk, that is, away from all confining walls, (ii) lowering the swimmer density so as to avoid particle clustering \cite{Navarro2015}, and (iii) choosing spherical active particles of small size, i.e., almost point-like, in order to reduce hydrodynamic backflow effects. However, the modulated activation gradients considered here certainly give rise to additional hydrodynamic contributions, of which the most prominent one is a self-polarization of the swimmer: the particle strives to align itself parallel or anti-parallel to the gradient, depending on its surface properties \cite{Bickel2014, Uspal2015a}. We addressed the influence of such a self-polarizing torque on the swimmer's diffusion in a recent study \cite{Geiseler2016SciRep} and concluded that for a small to moderate self-polarizing affinity, the tactic response of a swimmer behaves as reported in the present work. Its magnitude however slightly increases or decreases, subject to whether the swimmer tends to align itself parallel or anti-parallel to the gradient.

Finally, we remark that the setup considered in \ref{F1}(\textbf{a}) bears resemblance to that of Ref.\ \cite{Lozano2016}. However, a main difference between both setups is the way in which the spatial symmetry of the pulse waveform is broken, which was found to constitute the key factor---alongside the swimmer's finite persistence time---accountable for the emerge of any tactic drift. In the present setup, the pulse symmetry is broken due to the constant propagation of the pulses to the right, whereas in Ref.\ \cite{Lozano2016} an asymmetric pulse shape is considered. A tactic drift can be observed in both cases, however, the underlying mechanisms are rather different: in Ref.\ \cite{Lozano2016}, the observed tactic effect is explained with a saturation of the self-polarizing torque mentioned above, while in the model as considered in the present work, the swimmer's tactic drift solely results from the modulation of its active diffusion inside the traveling wave pulses.

\section{Results and Discussion}
\label{Results}


In the following we \emph{analytically} study the tactic drift of an artificial microswimmer subjected to traveling activation pulses. We assume that the spatio-temporal modulation of the swimmer's self-propulsion velocity has the form of a generic traveling wave, $v(x,t)=v_0 w[(x-ut)/L]$, with static profile $w(x/L)$. Upon changing coordinates from the resting laboratory frame to the co-moving wave frame, $x-ut\to x$, the Fokker-Planck equation (FPE) associated with the LEs \ (\ref{LE}) reads

\begin{equation}
\label{FPE}
\frac{\partial P(\mathbf{r},\phi,t)}{\partial t}=\left\{D_0\Delta-\boldsymbol{\nabla}\left[v_0 w\left(\frac{x}{L}\right)\mathbf{n}-\mathbf{u}\right]+D_\phi\frac{\partial^2}{\partial\phi^2}\right\}P(\mathbf{r},\phi,t),
\end{equation}
where $\mathbf{r}=(x,y)^\intercal$, $\mathbf{u}=(u,0)^\intercal$, $\mathbf{n}=(\cos\phi,\sin\phi)^\intercal$, and $\Delta$ and $\boldsymbol{\nabla}$ denote, respectively, the Laplace operator and the gradient in Cartesian coordinates $(x,y)$. The swimmer's dynamics perpendicular to the incoming wave exhibits no tactic behavior, since the pulse does not break the spatial symmetry in $y$ direction. Therefore, integrating over the $y$ coordinate and conveniently rescaling $x$ and $t$, $x=:Lx'$ and $t=:(L/v_0)t'$, we obtain a (still strictly Markovian) reduced FPE for the 2D marginal probability density $P(x',\phi,t')$, reading

\begin{equation}
\label{FPEscaled}
\frac{\partial P(x',\phi,t')}{\partial t'}=\left[\frac{D_0}{Lv_0}\frac{\partial^2}{\partial x'^2}-\frac{\partial}{\partial x'}\left(w(x')\cos\phi-\frac{u}{v_0}\right)+\frac{D_\phi L}{v_0}\frac{\partial^2}{\partial\phi^2}\right]P(x',\phi,t').
\end{equation}
Here, the effective rotational diffusion constant, $D_\phi L/v_0$, equals the ratio of the pulse width $L$ to the swimmer's persistence length $l_\phi=v_0/D_\phi$. The effective translational diffusion constant, $D_0/(Lv_0)$, corresponds instead to the ratio of the time the swimmer takes to ballistically travel a pulse width $L$ in a uniform activating medium, $L/v_0$, to the time it takes to diffuse the same length subject to the sole translational noise, $L^2/D_0$. This ratio characterizes the relative strength of translational fluctuations and coincides with the reciprocal of the P\'{e}clet number for mass transport. We agree now to drop the prime signs, so that in the remaining sections $x$ and $t$ denote the above dimensionless coordinates in the co-moving wave frame (unless stated otherwise).

\subsection{Diffusive Regime}
\label{DiffApprox}

The wave pulses can be wide and slow enough to regard the swimmer's motion inside each of them as purely diffusive. More precisely, this happens when the swimmer's rotational diffusion time, $D_\phi^{-1}$, is significantly smaller than the shortest ballistic pulse crossing time, $L/(v_0+u)$, i.e., when $D_\phi L/v_0\gg1+u/v_0$. Under this condition, we can further eliminate the orientational coordinate $\phi$, so that the effects of self-propulsion boil down to an effective 1D diffusive dynamics.
For this purpose, we apply to Eq.\ (\ref{FPEscaled}) the homogenization mapping procedure detailed in Ref.\ \cite{Kalinay2014} and obtain a partial differential equation for the marginal probability density

\begin{equation}
\label{fwdmap}
\mathcal{P}(x,t)=\int\limits_0^{2\pi}P(x,\phi,t)\,\mathrm{d}\phi.
\end{equation}
Following Ref.\ \cite{Geiseler2016EPJB}, we assume that the latter operation can be inverted by means of a ``backward'' operator $\hat{\psi}(x,\phi)$,

\begin{equation}
\label{bwdmap}
P(x,\phi,t)=\sum_{n=0}^\infty\epsilon^n\hat{\psi}_n(x,\phi)\frac{\mathcal{P}(x,t)}{2\pi},
\end{equation}
where $\hat{\psi}_0(x,\phi)=1$ and $\epsilon:=v_0/(D_\phi L)$. The expansion of $\hat{\psi}(x,\phi)$ in Eq.\ (\ref{bwdmap}) is justified by the fact that for $\epsilon\to0$ the swimmer rotates infinitely fast, in which case the self-propulsion can no longer contribute to its translational dynamics: the active particle behaves like a passive one, i.e., the rotational and translational dynamics decouple, and $P(x,\phi,t)$ simply becomes $\mathcal{P}(x,t)/(2\pi)$. Making use of Eqs.\ (\ref{fwdmap}) and (\ref{bwdmap}), respectively, in Eq.\ (\ref{FPEscaled}) and reordering all terms thus obtained according to their powers of $\epsilon$ \cite{Geiseler2016EPJB, Kalinay2014} yields a recurrence relation for the operators $\hat \psi_n$,

\begin{alignat}{2}
\label{Recursion}
\partial_\phi^2\hat{\psi}_{n+1}(x,\phi)=&\left[\hat{\psi}_n(x,\phi),\left(\frac{D_0}{Lv_0}\partial_x^2+\frac{u}{v_0}\partial_x\right)\right]+\cos\phi\,\partial_x w(x)\hat{\psi}_n(x,\phi)\nonumber\\
-&\frac{1}{2\pi}\sum_{m=0}^n\hat{\psi}_{n-m}(x,\phi)\partial_x w(x)\int\limits_0^{2\pi}\cos\phi\,\hat{\psi}_m(x,\phi)\,\mathrm{d}\phi,
\end{alignat}
where $[\custdots]$ denotes a commutator. By using the aforementioned initial condition $\hat{\psi}_0(x,\phi)=1$, the periodicity condition $\hat{\psi}_n(x,0)=\hat{\psi}_n(x,2\pi)$, and the normalization condition $\int_0^{2\pi}\hat{\psi}_n(x,\phi)\,\mathrm{d}\phi=2\pi\delta_{n,0}$, Eq.\ (\ref{Recursion}) can be solved iteratively, at least in principle, up to any arbitrarily high order. However, with increasing $n$ this task becomes more and more laborious and the results for the $\hat{\psi}_n$ read increasingly complicated.
In the diffusive limit however, the swimmer's rotational dynamics is significantly faster than its translational dynamics and $P(x,\phi,t)$ relaxes very fast in $\phi$ direction, that is, it only slightly differs from $\mathcal{P}(x,t)/(2\pi)$. It thus suffices to collect the terms of Eq.\ (\ref{bwdmap}) up to $\mathcal{O}(\epsilon)$, that is,

\begin{equation}
\label{bwdmapResult}
P(x,\phi,t)=\frac{1}{2\pi}[1-\epsilon\cos\phi\,\partial_x w(x)]\mathcal{P}(x,t).
\end{equation}
Finally, upon inserting Eq.\ (\ref{bwdmapResult}) into Eq.\ (\ref{FPEscaled}) and successively integrating with respect to $\phi$, we obtain the reduced 1D FPE \cite{Geiseler2016PRE}

\begin{equation}
\label{FPEmapped}
\frac{\partial\mathcal{P}(x,t)}{\partial t}=\hat{\mathbb{F}}(x)\mathcal{P}(x,t)=\left[\frac{\partial^2}{\partial x^2}\left(\frac{v_0}{2D_\phi L}\,w^2(x)+\frac{D_0}{Lv_0}\right)-\frac{\partial}{\partial x}\left(\frac{v_0}{4D_\phi L}\frac{\mathrm{d}w^2(x)}{\mathrm{d}x}-\frac{u}{v_0}\right)\right]\mathcal{P}(x,t),
\end{equation}
which describes the probability density of the swimmer's longitudinal position in the diffusive regime. Here, $\hat{\mathbb{F}}(x)$ denotes the Fokker-Planck operator, detailed on the right-hand side.

\subsubsection{Single Activation Pulse}
\label{SinglePulseDiff}

Following the presentation of Sec.\ \ref{Summary}, we first consider a single activating pulse hitting the swimmer and neglect translational fluctuations, $D_0=0$. The particle's tactic shift is then obtained by measuring its displacement from an initial position $x_0$, placed outside the pulse, on the right. Transforming back to the laboratory frame and taking the ensemble average, we define the tactic shift as $\Delta=:\langle x(t)-x_0+ut/v_0\rangle$. Note that $\Delta$ is still expressed in terms of the dimensionless units introduced above. We now can quantify the tactic shift in two ways: we either set a time $t$ and calculate the corresponding average swimmer's displacement in the pulse frame, $\langle x(t) \rangle$, hence

\begin{equation}
\label{Delta1}
\Delta(t)=\langle x(t)\rangle-x_0+\frac{u}{v_0}t,
\end{equation}
or, vice versa, we set the longitudinal shift, $x_1-x_0$, in the moving frame and calculate the corresponding \emph{mean first-passage time} $\langle t(x_1|x_0)\rangle$, hence

\begin{equation}
\label{Delta2}
\tilde{\Delta}(x_1)=x_1-x_0+\frac{u}{v_0}\langle t(x_1|x_0)\rangle.
\end{equation}
We remind that $\langle t(x_1|x_0)\rangle$ denotes the average time the particle takes to reach $x_1$ for the first time from $x_0$ \cite{Redner2001}.

As long as $x_1<x_0$, both methods are valid and equivalent, since in the moving frame the swimmer travels to the left and its position eventually takes on all values with $x<x_0$. However, for finite $t$ and $x_1$ we a priori do not know how to choose the values $x_1$ and $t$ that verify the identity $\Delta(t)=\tilde{\Delta}(x_1)$. However, if we consider the full shift of the swimmer after it has completely crossed the pulse (that is, for large enough $t$ or for $x_1$ placed far enough to the left of the pulse), both expressions yield the same result, that is, $\Delta(\infty)=\tilde{\Delta}(-\infty)$. This identity proved very helpful, since for the problem at hand the mean first-passage time can be calculated in a much simpler way than the average particle position. If the Fokker-Planck operator is time-independent, the mean first-passage time is the solution of the ordinary differential equation $\hat{\mathbb{F}}^\dagger(x)\langle t(x_1|x)\rangle=-1$ \cite{Goel1974, Risken1989}. Here, $\hat{\mathbb{F}}^\dagger$ is the adjoint Fokker-Planck operator acting upon the swimmer's starting position $x$, now taken as a variable, and $\langle t(x_1|x)\rangle$ obeys an absorbing boundary condition, $\langle t(x_1|x_1)\rangle=0$, at $x=x_1$. We thus have to solve the ordinary differential equation

\begin{equation}
\label{MFPTEq}
-1=\left[\frac{v_0}{2D_\phi L}w^2(x)\frac{\partial^2}{\partial x^2}+\left(\frac{v_0}{4D_\phi L}\frac{\mathrm{d}w^2(x)}{\mathrm{d}x}-\frac{u}{v_0}\right)\frac{\partial}{\partial x}\right]\langle t(x_1|x)\rangle.
\end{equation}
A second boundary condition follows naturally from the observation that outside of the pulse the swimmer's motion is deterministic. Namely, we know that $\dot{x}=-u/v_0$ at $x=x_0$, hence

\begin{equation}
\label{BC2}
\left.\frac{\partial\langle t(x_1|x)\rangle}{\partial x}\right|_{x=x_0}=\frac{v_0}{u}
\end{equation}
[because the swimmer starts at a position with $x>x_1$, to the right of the pulse, and crosses it to the left, increasing $x$ causes an increase in $\langle t(x_1|x)\rangle$].
With the above boundary conditions, Eq.\ (\ref{MFPTEq}) returns a unique solution,

\begin{equation}
\label{MFPT}
\langle t(x_1|x)\rangle=\int\limits_{x_1}^{x}\left[\frac{v_0}{u}\exp\left(\int\limits_y^{x_0} f(q)\,\mathrm{d}q\right)+\int\limits_y^{x_0}\frac{2}{\epsilon w^2(z)}\exp\left(\int\limits_y^z f(q)\,\mathrm{d}q\right)\mathrm{d}z\right]\mathrm{d}y,
\end{equation}
where $\epsilon=v_0/(D_\phi L)$ and

\begin{equation}
\label{f}
f(q):=\left(\frac{\epsilon}{4}\frac{\mathrm{d} w^2(q)}{\mathrm{d} q}-\frac{u}{v_0}\right)\left(\frac{\epsilon}{2}\,w^2(q)\right)^{-1}=\frac{\mathrm{d}\ln(q)}{\mathrm{d}q}-\frac{2u}{\epsilon v_0 w^2(q)}.
\end{equation}

\begin{figure}[htbp]
\centering
\includegraphics[width=0.94\textwidth]{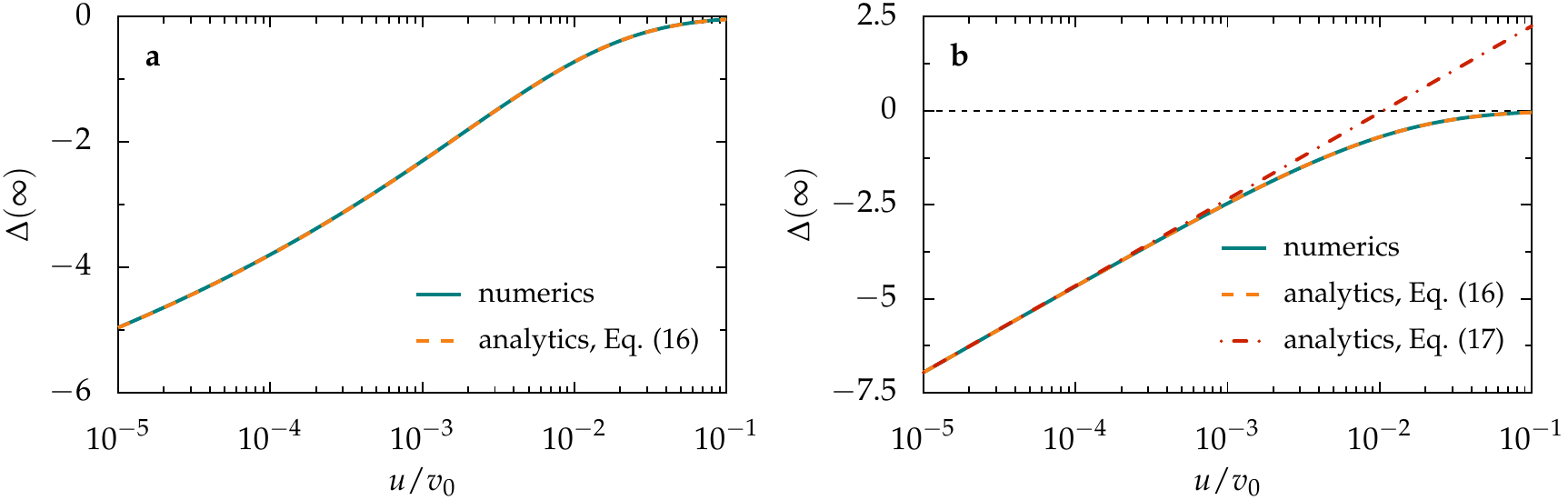}
\caption{Tactic shift of an artificial microswimmer across a single traveling pulse of the form (\textbf{a}) $w(x)=\exp(-x^2/2)$ and (\textbf{b}) $w(x)=\sech(x)$: $\Delta(\infty)$ vs. $u$ in units of the self-propulsion speed, $v_0$. The swimmer parameters are as in Fig.\ \ref{F1}(\textbf{b}): $v_0=53\,\mu\mathrm{m}/s$, $D_\phi=165\,\mathrm{s}^{-1}$, and $D_0=0$. We remind that here $x$ and $\Delta(\infty)$ are expressed in units of $L$. In (\textbf{a}), $L=4\,\mu\mathrm{m}$, i.e., about 12 times $l_\phi$; in (\textbf{b}), $L$ was set to $3.58\,\mu\mathrm{m}$, so that the two pulse profiles have the same half-width. The numerical results were obtained by solving the FPE (\ref{FPEscaled}).\label{F2}}
\end{figure}

For a smoothly decaying pulse profile $w(x)$, the condition for the swimmer to sweep through the entire pulse requires taking the limits $x_0 \to \infty$ and $x_1 \to -\infty$. The tactic shift of a swimmer in the diffusive regime is thus given by the expression

\begin{equation}
\label{Delta3}
\Delta(\infty)=\lim\limits_{\subalign{x_1&\to-\infty\\x_0&\to\infty}}\left\{x_1-x_0+\frac{u}{v_0}\int\limits_{x_1}^{x_0}\left[\frac{v_0}{u}\exp\left(\int\limits_y^{x_0} f(q)\,\mathrm{d}q\right)+\int\limits_y^{x_0}\frac{2}{\epsilon w^2(z)}\exp\left(\int\limits_y^z f(q)\,\mathrm{d}q\right)\mathrm{d}z\right]\mathrm{d}y\right\}.
\end{equation}
The r.h.s.\ of Eq.\ (\ref{Delta3}) contains two removable singularities; a partial integration yields the more compact result

\begin{equation}
\label{Delta4}
\Delta(\infty)=\int\limits_{-\infty}^\infty\frac{1}{w(y)}\int\limits_y^\infty\frac{\mathrm{d}w(z)}{\mathrm{d}z}\exp\left(-2\frac{D_\phi L}{v_0}\frac{u}{v_0}\int\limits_y^z\frac{1}{w^2(q)}\,\mathrm{d}q\right)\mathrm{d}z\,\mathrm{d}y.
\end{equation}
Note that this expression is independent of the boundary condition (\ref{BC2}). Indeed, outside the pulse, i.e., when $w(x)=0$, Eq.\ (\ref{MFPTEq}) reduces to a first-order differential equation and thus the boundary condition at $x=x_0$ becomes superfluous.
A comparison between the analytical prediction of Eq.\ (\ref{Delta4}) and results obtained by numerically integrating the FPE (\ref{FPEscaled}) is plotted in Fig.\ \ref{F2}. As in Sec.\ \ref{Summary}, for the activating pulse we chose a Gaussian profile, $w(x)=\exp(-x^2/2)$, of width $L\sim 12 l_\phi$. [We remark that due to the dimensionless scaling introduced at the beginning of this section, $L$ does not explicitly enter the waveform anymore, but instead it is incorporated into the effective diffusion constants, see Eq.\ (\ref{FPEscaled}).] The analytical and numerical curves for $\Delta(\infty)$ versus $u$ overlap in the regime of slow pulse speeds, $u\ll v_0$, thus confirming the validity of the diffusive approximation.

Moreover, for a soliton-like pulse profile, that is, $w(x)=\sech(x)$, we succeeded to obtain an explicit analytical expression for $\Delta(\infty)$, namely (see Appendix)

\begin{equation}
\label{DeltaSech}
\Delta(\infty)=\frac{\pi}{2}+\gamma-\ln\left(\frac{v_0}{D_\phi L}\frac{v_0}{u}\right),
\end{equation}
where $\gamma$ denotes the Euler-Mascheroni constant. Here, the agreement between numerical results and analytic approximation is quite close as well. The range of validity of Eq.\ (\ref{DeltaSech}) however shrinks to lower values of $u/v_0$, compared to the general result of Eq.\ (\ref{Delta4}), which is due to the fact that in the derivation of Eq.\ (\ref{DeltaSech}) we repeatedly assumed a very slow pulse propagation, see Eq.\ (\ref{FPEpiecewise}).

The analytical estimate of $\Delta(\infty)$ in Eq.\ (\ref{DeltaSech}) lends itself to a simple heuristic interpretation.
As mentioned in Sec.\ \ref{Summary}, in the diffusive regime the effective pulse half-width, $x_u$, is defined by the identity $w(x_u)=u/v_0$. Since for $u\ll v_0$ the swimmer propels itself inside an almost static pulse until it exits for good to its left, its tactic shift must be of the order of $x_u$. For the soliton-like profile $w(x)=\sech(x)$, this implies that

\begin{equation}
\label{DeltaRough}
\Delta(\infty)\approx -\ln\left(\frac{2v_0}{u}\right).
\end{equation}
Of course, this argument cannot fully reproduce Eq.\ (\ref{DeltaSech}). Nevertheless, it explains why the swimmer's tactic shift diverges in the limit $u\to 0$: as the pulse nearly comes to rest, its effective width grows exceedingly large; in the diffusive regime, the effect of the pulse's fore-rear symmetry breaking is therefore steadily enhanced.

Analogously, for the slow Gaussian pulse of Figs.\ \ref{F1}(\textbf{b}) and \ref{F2}(\textbf{a}), the dependence of $\Delta(\infty)$ on $u$ is expected to be of the form $\sqrt{2\ln(v_0/u)}$, also in good agreement with our numerical and analytical curves. Here, the pulse tails decay faster than for the soliton-like pulse, thus leading to a smaller tactic shift in the limit $u\to 0$.

\emph{The influence of translational noise.} We next consider the more realistic case with non-zero translational fluctuations, $D_0>0$. A very low translational noise level may be negligible in an appropriate range of pulse speeds. However, for $u\to 0$, the timescale on which the tactic shift approaches its asymptotic value, $\Delta (\infty)$, grows exceedingly long, which implies that at least in this regime, translational fluctuations must be taken into account. To a good approximation, the translational noise strength is independent of the spatio-temporal modulation of the swimmer's activation mechanism [see Eqs.\ (\ref{LE})]. As a main difference with the noiseless case $D_0=0$, in the presence of translational noise, the pulse edges are ``open'', as the swimmer can now cross them repeatedly back and forth. However, for sufficiently long observation times, the swimmer surely moves past the pulse, no matter how small $u$ and large $D_0$. Therefore, for $D_0>0$ we can calculate $\Delta(\infty)$ following the procedure already adopted for $D_0=0$. Even the boundary condition (\ref{BC2}) remains unchanged (and here is not superfluous!), since at $x=x_0$, that is, outside the pulse, we have $\langle\dot{x}\rangle=-u/v_0$. We thus obtain

\begin{equation}
\label{DeltaD0}
\Delta(\infty)=\int\limits_{-\infty}^\infty\int\limits_y^\infty\frac{w(z)\frac{\mathrm{d}w(z)}{\mathrm{d}z}}{\sqrt{\left[w^2(y)+\alpha\right]\left[w^2(z)+\alpha\right]}}\exp\left(-2\frac{D_\phi L}{v_0}\frac{u}{v_0}\int\limits_y^z\frac{1}{w^2(q)+\alpha}\,\mathrm{d}q\right)\mathrm{d}z\,\mathrm{d}y,
\end{equation}
with $\alpha:=2D_0 D_\phi/v_0^2$. Obviously, in the limit $D_0\to 0$ we recover Eq.\ (\ref{Delta4}).

\begin{figure}[htbp]
\centering
\includegraphics[width=0.45\textwidth]{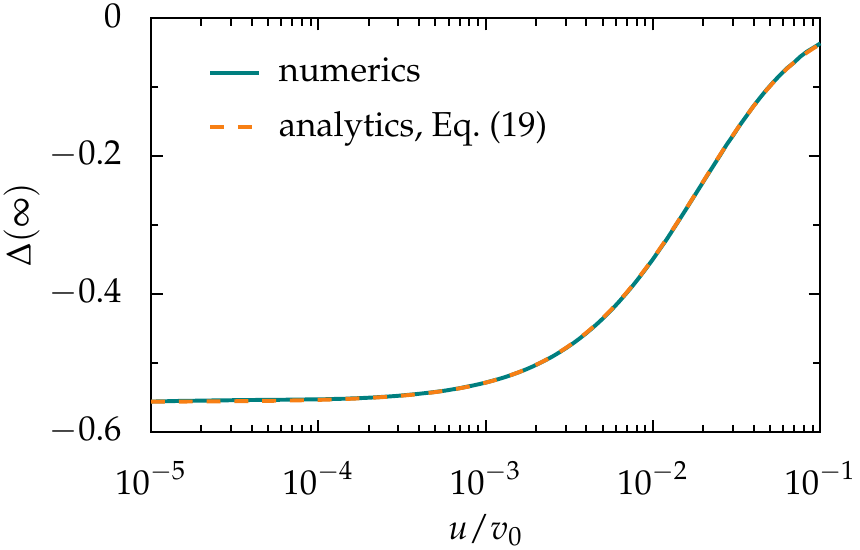}
\caption{Tactic shift $\Delta(\infty)$ as in Fig.\ \ref{F2} (\textbf{a}), but for non-zero translational noise with $D_0=2.2\,\mu\mathrm{m}^2/\mathrm{s}$. The numerical results were again obtained by solving the FPE (\ref{FPEscaled}).\label{F3}}
\end{figure}

In Fig.\ \ref{F3}, the dependence of $\Delta(\infty)$ on the pulse speed $u$ was determined both by computing the integrals in Eq.\ (\ref{DeltaD0}) and numerically solving Eq.\ (\ref{FPEscaled}). Again, the agreement between analytical and numerical results is quite close. We notice that in the presence of translational noise, the limit of $\Delta(\infty)$ for $u\to0$ is finite. We attribute this property to the fact that translational diffusion, which tends to suppress tactic rectification, prevails over self-propulsion, but only in the pulse's tails. More precisely, the swimmer's dynamics is dominated by translational diffusion when $D_0/(Lv_0)\gg v_0w^2(x)/(2D_\phi L)$ [see Eq.\ (\ref{FPEmapped})], or $w^2(x)\ll\alpha$ [see Eq.\ (\ref{DeltaD0})]. Under this condition, a natural definition of the effective pulse width is ${\rm min} [x_u,x_t]$, with $x_t$ being the solution of the equation $w(x_t)\propto\sqrt{2D_0D_\phi/v_0^2}$. On decreasing $u$, the ratio $x_u/x_t$ diverges, the effective pulse width coincides with $x_t$, and $\Delta(\infty)$ becomes a function of the sole parameter $\alpha=2D_0D_\phi/v_0^2$.

\subsubsection{Periodic Pulse Train}
\label{PulseTrainDiff}

In the following, we consider a periodic sequence of activating pulses, $w(x+n)=w(x)\ \forall n\in\mathbb{Z}$, with unit period (in dimensionless units), which corresponds to a period of $L$ in the unscaled notation of Sec.\ \ref{Summary}. To calculate the resulting longitudinal drift speed, $v_x=\lim_{t\to\infty}\langle\dot{x}\rangle+u/v_0$ from Eq.\ (\ref{FPEmapped}), we introduce the reduced one-zone probability

\begin{equation}
\label{OneZone}
\tilde{\mathcal{P}}(x,t)=\sum\limits_{n=-\infty}^\infty\mathcal{P}(x+n,t),
\end{equation}
which maps the overall probability density $\mathcal{P}(x,t)$ onto one period of the pulse sequence \cite{Burada2007, Burada2009, Geiseler2016PRE}. That is, instead of considering the time-evolution of the swimmer's probability density along an infinite periodic pulse sequence, we focus on a single wave period and impose periodic boundary conditions to ensure the existence of a stationary state.
Accordingly, we define the corresponding reduced probability current, $\tilde{J}(x,t)$, and obtain the continuity equation

\begin{equation}
\label{ContEq}
\frac{\partial\tilde{\mathcal{P}}(x,t)}{\partial t}=-\frac{\partial\tilde{J}(x,t)}{\partial x},
\end{equation}
where, upon introducing the two auxiliary functions $g(x):=v_0w^2(x)/(2D_\phi L)+D_0/(Lv_0)$ and $h(x):=v_0(\mathrm{d}/\mathrm{d}x)w^2(x)/(4D_\phi L)-u/v_0$, $\tilde{J}(x,t)$ can be written in a compact form as

\begin{equation}
\label{J}
\tilde{J}(x,t)=-\exp\left(\int\limits_0^x\frac{h(y)}{g(y)}\,\mathrm{d}y\right)\frac{\partial}{\partial x}\,g(x)\exp\left(-\int\limits_0^x\frac{h(y)}{g(y)}\,\mathrm{d}y\right)\tilde{\mathcal{P}}(x,t).
\end{equation}
In the stationary limit, $\tilde{J}(x,t\to \infty)=:\tilde{J}_{\mathrm{st}}$ becomes constant and can be calculated explicitly \cite{Burada2009},

\begin{alignat}{2}
\label{JResult}
\tilde{J}_{\mathrm{st}}&=\frac{v_0}{2D\phi L}\left[1-\exp\left(2\frac{D_\phi L}{v_0}\frac{u}{v_0}\int\limits_0^1\frac{1}{w^2(x)+\alpha}\,\mathrm{d}x\right)\right]\nonumber\\
&\times\left[\int\limits_0^1\int\limits_0^1\frac{1}{\sqrt{\left[w^2(x)+\alpha\right]\left[w^2(x+y)+\alpha\right]}}\exp\left(2\frac{D_\phi L}{v_0}\frac{u}{v_0}\int\limits_x^{x+y}\frac{1}{w^2(z)+\alpha}\,\mathrm{d}z\right)\mathrm{d}y\,\mathrm{d}x\right]^{-1}.
\end{alignat}
%
Upon transforming back to the laboratory frame, we finally obtain a simple expression for the swimmer's tactic drift speed, namely

\begin{equation}
\label{vxDiff}
v_x=\int\limits_0^1\tilde{J}_\mathrm{st}\,\mathrm{d}x+\frac{u}{v_0}=\tilde{J}_\mathrm{st}+\frac{u}{v_0}.
\end{equation}

Taking the limit $D_0\to 0$ and assuming $w(0)=w(1)=0$ [as for $w(x)=\sin^2(\pi x)$ in Figs.\ \ref{F1}(\textbf{c,d})], Eq.\ (\ref{JResult}) can be given the more convenient form

\begin{equation}
\label{JResultNoD0}
\tilde{J}_{\mathrm{st}}=\frac{u}{v_0}\left[\int\limits_0^1\int\limits_0^1\frac{w'(x+y)}{w(x)}\exp\left(-2\frac{D_\phi L}{v_0}\frac{u}{v_0}\int\limits_{x+y}^{x+1}\frac{1}{w^2(z)}\,\mathrm{d}z\right)\mathrm{d}y\,\mathrm{d}x\right]^{-1},
\end{equation}
where the prime sign denotes the derivative with respect to the function's argument.

\begin{figure}[htbp]
\centering
\includegraphics[width=0.94\textwidth]{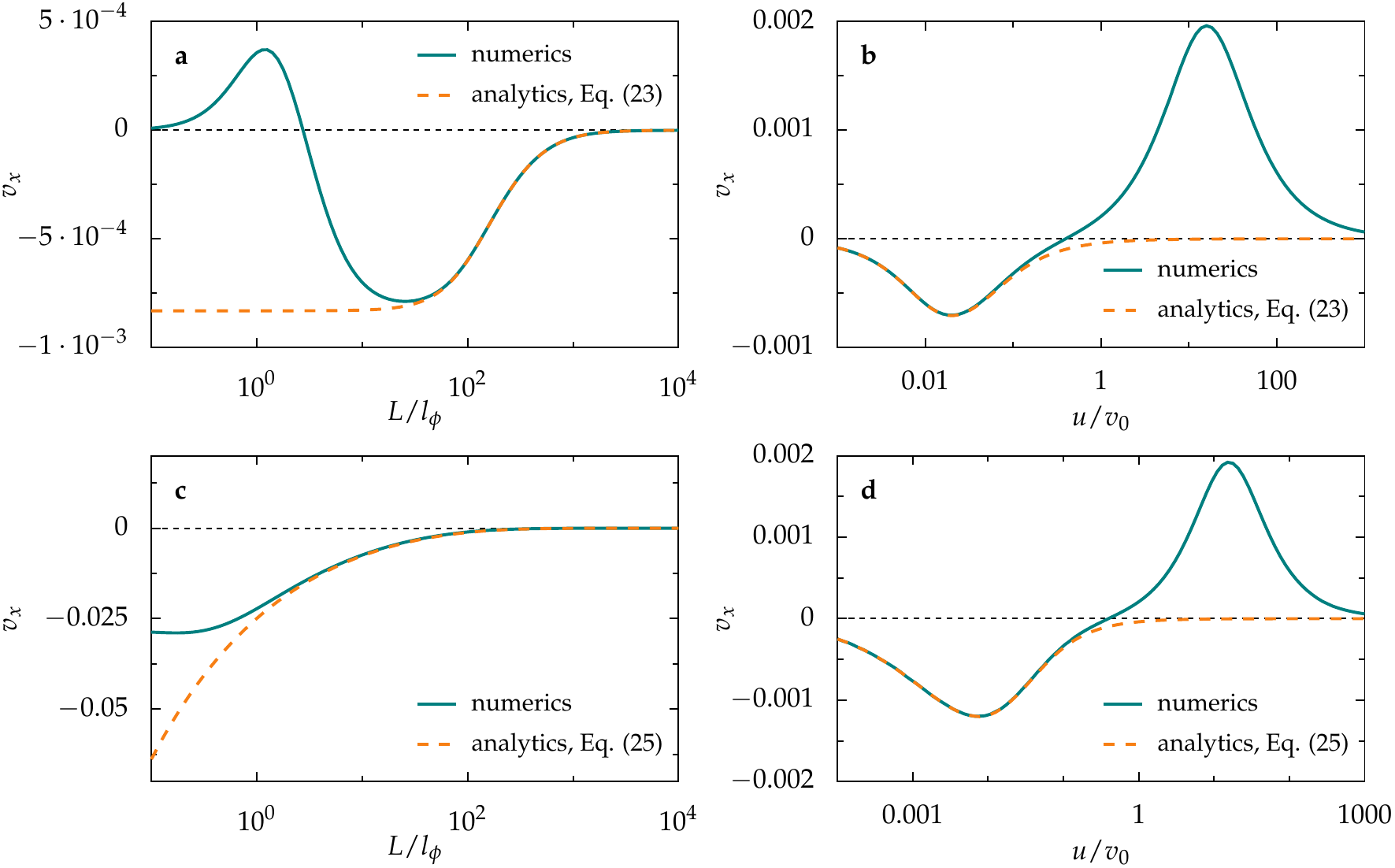}
\caption{Tactic drift velocity of an artificial microswimmer subjected to the sinusoidal activating pulse sequence of Figs.\ \ref{F1}(\textbf{c,d}): $v_x$ vs.\ the pulse width $L$ (\textbf{a,c}) and pulse speed $u$ (\textbf{b,d}). The swimmer parameters $v_0$ and $D_\phi$ are the same as in the previous figures and $D_0=2.2\,\mu\mathrm{m}^2/\mathrm{s}$ in (\textbf{a,b}) [$D_0=0$ in (\textbf{c,d})]. Furthermore, $u=0.01v_0$ in (\textbf{a,c}) and $L=100l_\phi$ in (\textbf{b,d}). The numerical data plotted here have been obtained by numerically integrating the LEs (\ref{LE}) or, equivalently, the FPE (\ref{FPEscaled}).\label{F4}}
\end{figure}

In Fig.\ \ref{F4} we compare the analytical approximation of Eqs.\ (\ref{JResult}-\ref{JResultNoD0}) with the exact values for $v_x$, computed by numerically integrating the FPE (\ref{FPEscaled}) or the LEs (\ref{LE}). As to that, we remark that both numerical approaches yield---within their accuracy---the same results, so that we can adopt either of them, as more convenient. In general, solving the FPE is advantageous, since numerically integrating the LEs for an ensemble of particles is rather time-consuming. For some parameter ranges however, namely when the probability density $P(x,\phi,t)$ is sharply peaked, the spatial grid, on which the temporal evolution of the FPE is solved, has to be extremely fine. Memory consumption and computation time then explode, so that numerically integrating the LEs proves more effective.

As expected, a close agreement between the numerical and analytical curves in Fig.\ \ref{F4} is achieved if both conditions $D_\phi L/v_0\gg1$ and $u/v_0\ll1$ are {\it simultaneously} fulfilled. In contrast to our initial conjecture, under the weaker condition $D_\phi L/v_0\gg1+u/v_0$, the motion of an active swimmer inside a traveling pulse may well be regarded as purely diffusive, but the corresponding diffusive approximation fails to correctly predict its tactic drift when $u\gtrsim v_0$ [see Figs.\ \ref{F4}(\textbf{b},\textbf{d})].

As a consequence, we find that the {\it positive} branches of the $v_x$ curves are purely determined by the ballistic nature of the swimmer's dynamics (which is indeed rather subordinate for $L\gg l_\phi$, but nevertheless cannot be neglected if $u\gtrsim v_0$). This conclusion is supported by Figs.\ \ref{F4}(\textbf{a},\textbf{c}), where for $L/l_\phi \lesssim 10$ the tactic response clearly depends on $D_0$ and, more importantly, analytic and numerical curves seem to part ways. In the diffusive approximation (dashed curves), the effect of the translational fluctuations is predicted to just prevent the drift from growing more negative, whereas in the full dynamics treatment (solid curves) the influence of $D_0$ causes $v_x$ to change sign.

\subsection{Ballistic Regime}
\label{BallSec}

We focus now on the opposite dynamical regime, termed ballistic. Here, the traveling pulses are assumed to be so narrow and sweep through the swimmer so fast that the swimmer's orientation almost does not change during a single pulse crossing, i.e., the time a single activating pulse takes to pass the swimmer is negligible with respect to the angular diffusion time $D_\phi^{-1}$. In such a limit we take $\phi$ constant and rewrite the FPE (\ref{FPEscaled}) as

\begin{equation}
\label{FPEfixed}
\frac{\partial P_\phi(x,t)}{\partial t}=\left[\frac{D_0}{Lv_0}\frac{\partial^2}{\partial x^2}-\frac{\partial}{\partial x}\left(w(x)\cos\phi-\frac{u}{v_0}\right)\right]P_\phi(x,t),
\end{equation}
where $P_\phi(x,t)$ is the corresponding conditional probability density at fixed angle $\phi$. In the following, we make use of Eq.\ (\ref{FPEfixed}) to calculate the conditional tactic shift, $\Delta_\phi(\infty)$, or drift, $v^\phi_x$, as appropriate. Since the angular coordinate is actually not fixed, but rather freely diffusing on an exceedingly long timescale, the quantities $\Delta_\phi(\infty)$ and $v^\phi_x$ will be eventually averaged with respect to $\phi$, which is uniformly distributed on the interval $[0,2\pi]$.

\subsubsection{Single Activation Pulse}
\label{SinglePulseBall}

The tactic shift $\Delta(\infty)$ of a swimmer swept through a single activating pulse can be calculated, once again, as in Sec.\ \ref{SinglePulseDiff}, namely

\begin{equation}
\label{DeltaBallisticD0}
\Delta(\infty)=\frac{1}{2\pi}\int\limits_0^{2\pi}\Delta_\phi(\infty)\,\mathrm{d}\phi=\frac{Lv_0}{D_0}\int\limits_{-\infty}^\infty\int\limits_x^\infty w(y)\exp\left[-\frac{Lu}{D_0}(y-x)\right]I_1\left(\frac{Lv_0}{D_0}\int\limits_x^yw(z)\,\mathrm{d}z\right)\mathrm{d}y\,\mathrm{d}x,
\end{equation}
where $I_1(x):=(1/\pi)\int_0^\pi\exp(x\cos\phi)\cos\phi\,\mathrm{d}\phi$ is a modified Bessel function of the first kind \cite{Abramowitz1965}.
Although in the absence of translational fluctuations, $D_0=0$, the swimmer's fixed-angle dynamics is purely deterministic, we can still employ the mean first-passage time technique to calculate $\Delta(\infty)$ for $D_0=0$, yielding

\begin{equation}
\label{DeltaBallistic}
\Delta(\infty)=\int\limits_{-\infty}^\infty\left(\frac{1}{\sqrt{1-\left(v_0^2/u^2\right)w^2(x)}}-1\right)\mathrm{d}x,
\end{equation}
which surely is well-defined in the ballistic regime with $u>v_0$. Here, the positive tactic shift must be attributed to the fact that swimmers oriented to the right, i.e., parallel to the direction of pulse propagation, ``surf'' the pulse for a longer time than swimmers oriented in the opposite direction.

\begin{figure}[htbp]
\centering
\includegraphics[width=0.94\textwidth]{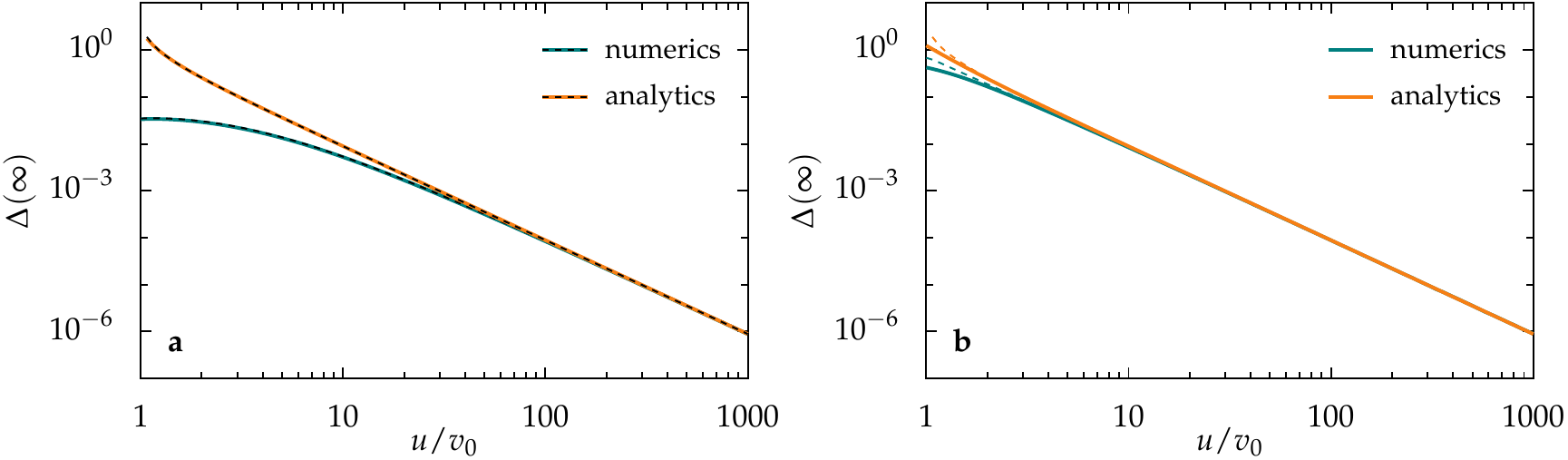}
\caption{Tactic shift of an artificial microswimmer hit by a single Gaussian pulse, like in Figs.\ \ref{F2} and \ref{F3}, but for larger values of the pulse speed $u$. The particle parameters are as in the previous figures; furthermore $L=1\,\mu\mathrm{m}$ ($\sim3l_\phi$) in (\textbf{a}) and $0.1\,\mu\mathrm{m}$ in (\textbf{b}). The numerical curves were obtained by solving the FPE (\ref{FPEscaled}) or integrating the LEs (\ref{LE}); the analytical curves were calculated in the ballistic approximation of Eqs.\ (\ref{DeltaBallisticD0}) or (\ref{DeltaBallistic}), as appropriate. For the sake of a comparison, we plotted the curves for $D_0=2.2\,\mu\mathrm{m}^2/\mathrm{s}$ (solid lines) together with the corresponding curves for $D_0=0$ (dashed lines).\label{F5}}
\end{figure}

By inspecting Fig.\ \ref{F5}, we notice that the ballistic approximation holds good for fast activating pulses. One might expect it to work well only if the swimmer's rotational diffusion time, $D_\phi^{-1}$, is larger than the timescale on which a swimmer oriented to the right ($\phi=0$) ballistically crosses the pulse, $L/(u-v_0)$. By analogy with Sec.\ \ref{DiffApprox}, one would end up with the condition $u/v_0\gg1+L/l_\phi$. This argument however totally disregards the influence of translational fluctuations and thus only applies when $D_0/(Lv_0)$ can be safely neglected [see Fig.\ \ref{F5}(\textbf{a})]. More in general, we must require that $D_\phi^{-1}$ is larger than the pulse crossing timescale in the ballistic regime, $L/(u-v_0)$, {\it or} in the diffusive regime, $L^2/D_0$, \emph{whichever is smaller}. This leads to the weaker condition for the validity of the ballistic approximation, $l_\phi/L\gg\min\big[(u/v_0-1)^{-1},Lv_0/D_0\big]$.

By comparing the data for $D_0=0$ and $D_0>0$ in Fig.\ \ref{F5}, we also observe that translational fluctuations affect the tactic response of a ballistic swimmer only marginally: contrary to the diffusive regime, here the swimmer crosses the pulse quite fast, so that the translational noise has almost no time to act on it (provided the pulses being not too narrow). The simple expression of Eq.\ (\ref{DeltaBallistic}) can thus be safely employed to predict the tactic shift of a swimmer in the ballistic regime also in the presence of translational noise.

\subsubsection{Periodic Pulse Train}
\label{PulseTrainBall}

For the periodic sequence of activation pulses introduced in in Sec.\ \ref{PulseTrainDiff}, the swimmer's tactic drift can also easily be calculated in the ballistic approximation and we obtain analogously as in Sec.\ \ref{PulseTrainDiff}

\begin{alignat}{2}
\label{vxBallD0}
v_x=\frac{D_0}{2\pi Lv_0}\int\limits_0^{2\pi}&\left\{1-\exp\left[\frac{Lv_0}{D_0}\left(\frac{u}{v_0}-\cos\phi\int\limits_0^1w(x)\,\mathrm{d}x\right)\right]\right\}\nonumber\\
\times&\left\{\int\limits_0^1\int\limits_0^1\exp\left[\frac{Lv_0}{D_0}\left(\frac{u}{v_0}\,y-\cos\phi\int\limits_x^{x+y}w(z)\,\mathrm{d}z\right)\right]\mathrm{d}y\,\mathrm{d}x\right\}^{-1}\mathrm{d}\phi+\frac{u}{v_0}.
\end{alignat}

If we further neglect translational fluctuations, $D_0=0$, in the ballistic regime the longitudinal LE (\ref{LE}) simplifies to a purely deterministic fixed-angle equation of motion, $\dot{x}=w(x)\cos\phi-u/v_0$. For a sinusoidal pulse sequence, $w(x)=\sin^2(\pi x)$, this equation can be solved analytically, i.e.,

\begin{equation}
\label{DetSol}
x(t)=\left\{\begin{array}{ll}
-\frac{1}{\pi}\arctan\left[\frac{\tan\left(\pi tu/v_0\sqrt{1-(v_0/u)\cos\phi}\right)}{\sqrt{1-(v_0/u)\cos\phi}}\right]&:t\leq\frac{v_0}{2u\sqrt{1-(v_0/u)\cos\phi}}\\
-\frac{1}{\pi}\arctan\left[\frac{\tan\left(\pi tu/v_0\sqrt{1-(v_0/u)\cos\phi}\right)}{\sqrt{1-(v_0/u)\cos\phi}}\right]-1&:t>\frac{v_0}{2u\sqrt{1-(v_0/u)\cos\phi}},
\end{array}\right.
\end{equation}
with $x$ restricted to the interval $[-1,0]$, $u>v_0$, and initial condition $x(0)=0$. The ballistic pulse crossing time $t_\mathrm{c}^\phi$ for a fixed orientation angle, defined by the relation $x(t_\mathrm{c}^\phi)=-1$, thus reads $t_\mathrm{c}^\phi=v_0/(u\sqrt{1-(v_0/u)\cos\phi})$. The swimmer's tactic drift can then be calculated using the known relation $v^\phi_x=-1/t_\mathrm{c}^\phi+u/v_0$. If $u$ grows smaller than $v_0$, however, particles oriented to the right can get trapped inside the pulses. This occurs when their self-propulsion speed in $x$ direction, $v_0w(x)\cos\phi$, compensates for the translational speed, $-u$. As $w(x)$ is valued between zero and one, swimmers get trapped with orientation $-\arccos(u/v_0)<\phi<\arccos(u/v_0)$. In the co-moving pulse frame, the velocity of trapped swimmers is zero, so that the $\phi$-averaged drift velocity turns out to be

\begin{equation}
\label{vxBall}
v_x=\left\{\begin{array}{ll}
-\frac{u}{v_0}\left(\frac{1}{2\pi}\int\limits_0^{2\pi}\sqrt{1-\frac{v_0}{u}\cos\phi}\,\mathrm{d}\phi-1\right)&:\frac{u}{v_0}\ge1\\
-\frac{u}{v_0}\left(\frac{1}{2\pi}\int\limits_{\arccos(u/v_0)}^{2\pi-\arccos(u/v_0)}\sqrt{1-\frac{v_0}{u}\cos\phi}\,\mathrm{d}\phi-1\right)&:\frac{u}{v_0}<1.
\end{array}\right.
\end{equation}

It is interesting to remark that we can now refine the validity criterion for the ballistic approximation discussed in the previous section, owing to the more precise estimate of the ballistic pulse-crossing time derived above. Following the relevant argument of Sec.\ \ref{SinglePulseBall}, we thus expect the ballistic approximation to hold for $l_\phi/L>\min\left(t_\mathrm{c}^{\phi=0},Lv_0/D_0\right)$.

\begin{figure}[htbp]
\centering
\includegraphics[width=0.94\textwidth]{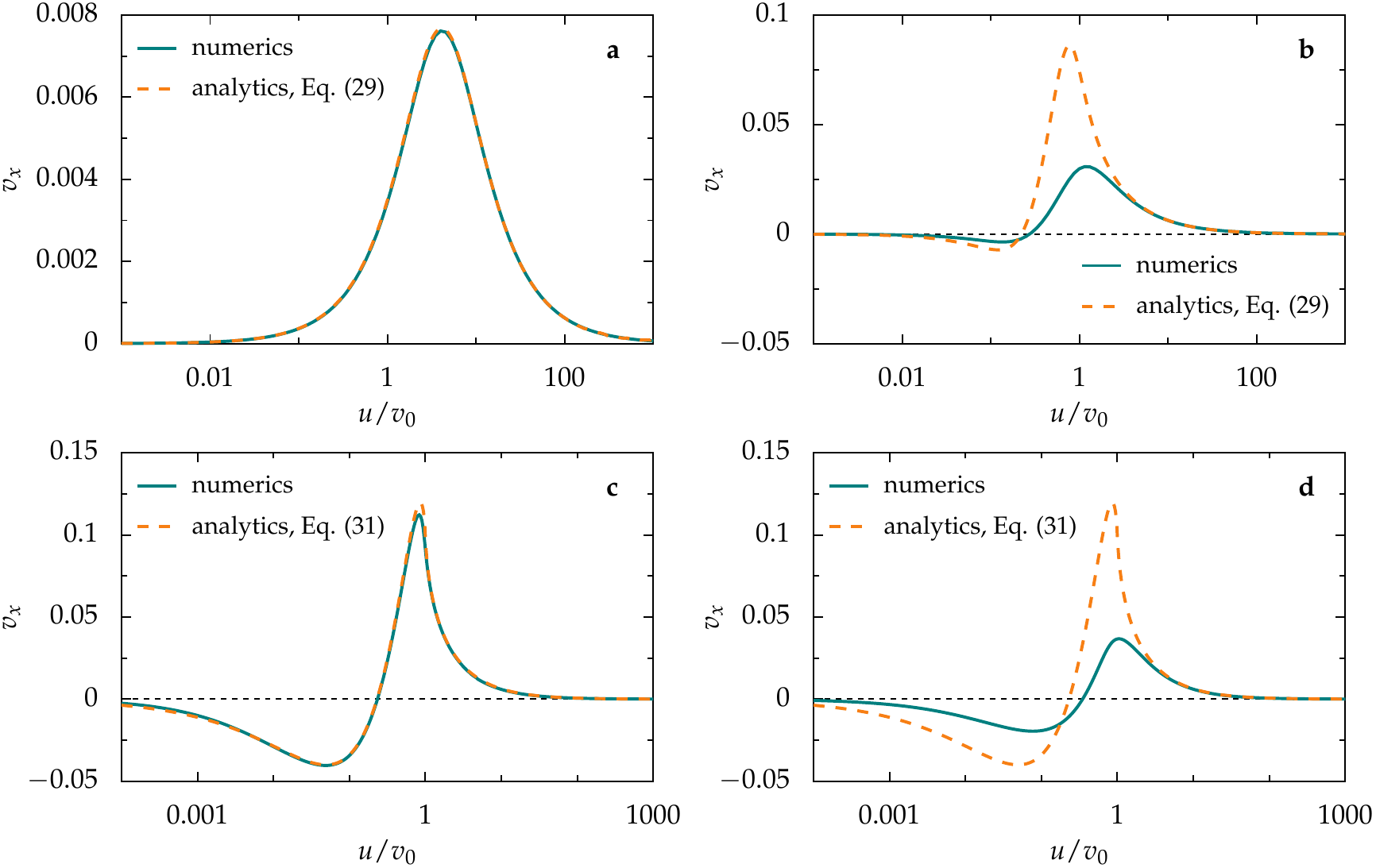}
\caption{Tactic drift of an artificial microswimmer induced by the sinusoidal pulse sequence of Fig.\ \ref{F4}: $v_x$ vs.\ $u$ in units of $v_0$. In (\textbf{a,c}) we chose a very small pulse periodicity, $L=0.2l_\phi$, whereas in (\textbf{b,d}) $L$ was set to $5l_\phi$. The swimmer parameters $v_0$ and $D_\phi$ were chosen as in the previous figures and set $D_0=2.2\,\mu\mathrm{m}^2/\mathrm{s}$ in (\textbf{a,b}) and $D_0=0$ in (\textbf{c,d}). The numerical curves were obtained by numerically integrating the LEs (\ref{LE}) or solving the FPE (\ref{FPEscaled}).\label{F6}}
\end{figure}

A comparison between exact numerics and the ballistic approximation is shown in Fig.\ \ref{F6}. As expected, its range of validity in the parameter $u$ shrinks on increasing $L/l_\phi$ and the refined validity condition just introduced provides an estimate of that range.
Another interesting property illustrated in Fig.\ \ref{F6} is that the ballistic approximation also predicts a regime of negative tactic drift, which we explain as follows. We have already mentioned that for $u>v_0$ all swimmers surely cross the wave pulse and the positive net drift results from the fact that particles oriented parallel to the direction of pulse propagation spend on average a longer time inside the pulse than particles oriented in the opposite direction. For $u<v_0$, however, the emerging trapping mechanism causes swimmers with $-\arccos(u/v_0)<\phi<\arccos(u/v_0)$ to travel to the right with velocity $u$. If $u$ is suitably smaller than $v_0$, the trapped swimmers may happen to move considerably slower to the right than swimmers with $|\phi|>\pi/2$ to the left, thus causing a negative net drift.

\section{Conclusions}
\label{resume}
In summary, we analytically showed that the dynamics of artificial microswimmers subjected to traveling activation pulses manifests two, partially competing tactic effects, both induced by the broken spatial symmetry associated with the pulse propagation. In the two limiting regimes of high and low rotational fluctuations, defined with respect to the pulse parameters $u$ and $L$, we obtained analytical approximations that are in close agreement with the exact numerical results. Likewise, these analytical results compare favorably with the numerical data reported before in Ref.\ \cite{Geiseler2016PRE}. Our analytical approach provides a valuable framework for future studies of the tactic response of artificial microswimmers in spatio-temporally modulated activation media. Moreover, we identified the positive tactic drift as being a purely ballistic effect, i.e., to stem solely from to the finite persistence of the swimmer's active Brownian motion, whereas the negative tactic drift results from the combination of diffusive and ballistic properties of the swimmer's dynamics.

A generalization of the single particle model considered in the present work to multiple interacting swimmers---slightly similar to the setup considered in Ref.\ \cite{Mijalkov2016} for macroscopic phototactic robots---could also give rise to interesting new collective effects, primarily stemming from the coupling of the hydrodynamic swimmer interactions to the hydrodynamic influence of the activation gradient \cite{Geiseler2016SciRep}.

\vspace{6pt}


\acknowledgments{This work has been supported by the cluster of excellence Nanosystems Initiative Munich (P.H.). P.H. and F.M. acknowledge a financial support from the Center for Innovative Technology (ACIT) of the University of Augsburg. F.M. also thanks the DAAD for a Visiting Professor grant.}

\authorcontributions{A.G. performed all calculations in this project. All authors contributed to the planning and the discussion of the results as well as to the writing of this work.}

\conflictofinterests{The authors declare no conflict of interest.}



\appendixtitles{yes} 
\appendixsections{one} 
\appendix
\section{Tactic Shift Induced by a Soliton-Like Pulse}
\label{App}

Let the activating pulse have a simple exponentially decaying profile, $w(x)=\sech(x)$, and $D_0=0$. Starting from Eq.\ (\ref{FPEmapped}), we further rescale the time, $t=(v_0/u)\tau$, which leaves only one effective parameter, $\eta:=v_0^2/(2D_\phi Lu)$, in the resulting FPE. Upon introducing the auxiliary coordinate $\chi$, $x=\arsinh(\sqrt{\eta}\,\chi)$, we rewrite the new FPE as

\begin{equation}
\label{FPESech}
\frac{\partial\mathcal{P}(\chi,\tau)}{\partial\tau}=\left(\frac{\partial^2}{\partial\chi^2}+\frac{\partial}{\partial\chi}\sqrt{\frac{1}{\eta}+\chi^2}\right)\mathcal{P}(\chi,\tau),
\end{equation}
which for slow wave pulses, $u\ll v_0^2/(2D_\phi L)$ or $\eta\gg1$, respectively, can be approximated by

\begin{equation}
\label{FPEpiecewise}
\frac{\partial\mathcal{P}(\chi,\tau)}{\partial\tau}=\left(\frac{\partial^2}{\partial\chi^2}+\frac{\partial}{\partial\chi}|\chi|\right)\mathcal{P}(\chi,\tau).
\end{equation}
For $\chi\geq0$ $[\chi<0]$, the corresponding Fokker-Planck operator is associated with an Hermitian operator, $\hat{\mathbb{F}}(\chi)\to\exp(\chi^2/4)\hat{\mathbb{F}}(\chi)\exp(-\chi^2/4)$ $\big[\hat{\mathbb{F}}(\chi)\to\exp(-\chi^2/4)\hat{\mathbb{F}}(\chi)\exp(\chi^2/4)\big]$ \cite{Risken1989}. Accordingly, the FPE (\ref{FPEpiecewise}) can be mapped onto the Schrödinger equation for a particle in the piecewise harmonic potential

\begin{displaymath}
V(\chi)=\left\{
\begin{array}{ll}
\vphantom{\Big(}\frac{1}{4}(\chi^2-2): & \chi\geq0\\
\vphantom{\Big(}\frac{1}{4}(\chi^2+2): & \chi<0.
\end{array}\right.
\end{displaymath}
In principle, the probability density $\mathcal{P}(\chi,\tau)$ could be expressed in terms of the eigenvalues and eigenfunctions of such Schrödinger equation, but in view of the potential cusp at $\chi=0$ that would be a challenging task.
Therefore, we again resort to computing the mean first-passage time, $\langle{\tau}(\chi_1|\chi)\rangle$, by solving the relevant differential equation associated with the FPE \ (\ref{FPEpiecewise}), namely

\begin{equation}
\label{MFPTEqPiecewise}
-1=\left(\frac{\partial^2}{\partial\chi^2}-|\chi|\frac{\partial}{\partial\chi}\right)\langle{\tau}(\chi_1|\chi)\rangle,
\end{equation}
with the boundary and continuity conditions

\begin{alignat}{2}
&\text{i)}\quad&&\langle{\tau}(\chi_1|\chi_1)\rangle=0,\nonumber\\
&\text{ii)}\quad&&\langle{\tau}(\chi_1|0^+)\rangle=\langle{\tau}(\chi_1|0^-)\rangle,\nonumber\\
&\text{iii)}\quad&&\left.\frac{\partial\langle{\tau}(\chi_1|\chi)\rangle}{\partial\chi}\right|
_{\chi=0^+}=\left.\frac{\partial\langle{\tau}(\chi_1|\chi)\rangle}{\partial\chi}\right|_{\chi=0^-},\nonumber\\
&\text{iv)}\quad&&\left.\frac{\partial\langle{\tau}(\chi_1|\chi)\rangle}{\partial\chi}\right|_{\chi\to\infty}=0.\nonumber
\end{alignat}
Its solution for $\chi\geq0$ reads

\begin{equation}
\label{MFPTPiecewise}
\langle{\tau}(\chi_1|\chi)\rangle=\frac{\pi}{2}\left[\erfi\left(\frac{\chi}{\sqrt{2}}\right)-\erf\left(\frac{\chi_1}{\sqrt{2}}\right)\right]
-\frac{\chi^2}{2}\,{}_2F_2\left(1,1;\frac{3}{2},2;\frac{\chi^2}{2}\right)+\frac{\chi_1^2}{2}\,{}_2F_2\left(1,1;\frac{3}{2},2;-\frac{\chi_1^2}{2}\right),
\end{equation}
where $\erfi(x)=2/\sqrt{\pi}\int_0^x\exp\left(v^2\right)\mathrm{d}v$ is the imaginary error function and

\begin{equation}
\label{2F2Int}
{}_2F_2\left(1,1;\frac{3}{2},2;x\right)=\frac{\sqrt{\pi}}{x}\int\limits_0^{\sqrt{x}}\erf(v)\exp\left(v^2\right)\mathrm{d}v
\end{equation}
is a generalized hypergeometric function \cite{Prudnikov1992}. The swimmer's tactic shift can now be formally computed as

\begin{alignat}{2}
\label{SechResult}
\Delta(\infty)=&\lim\limits_{\subalign{x_1&\to-\infty\\x_0&\to\infty}}\left\{x_1-x_0+\frac{\pi}{2}\left[\erfi\left(\frac{\sinh(x_0)}{\sqrt{2\eta}}\right)-\erf\left(\frac{\sinh(x_1)}{\sqrt{2\eta}}\right)\right]\right.\nonumber\\
&-\frac{\sinh^2(x_0)}{2\eta}\,{}_2F_2\left(1,1;\frac{3}{2},2;\frac{\sinh^2(x_0)}{2\eta}\right)+\left.\frac{\sinh^2(x_1)}{2\eta}\,{}_2F_2\left(1,1;\frac{3}{2},2;-\frac{\sinh^2(x_1)}{2\eta}\right)\right\}.
\end{alignat}
[We remind that in the present notation the particle displacement in the laboratory frame is calculated as $x(t)-x_0+\tau$).]
To explicitly take the above limits, one must determine the asymptotic expansions of the special functions in Eq.\ (\ref{SechResult}). For $\erfi(x)$, this can be easily accomplished \cite{Dingle1958},

\begin{equation}
\label{ErfiExpansion}
\erfi(x)\sim\frac{\exp\left(x^2\right)}{\sqrt{\pi}\,x}.
\end{equation}

The expansion of the hypergeometric function ${}_2F_2$ for $x \to \pm\infty$ is somewhat more elaborate. We start by considering its integral representation for {\it negative} arguments,

\begin{equation}
\label{2F2IntNeg}
{}_2F_2\left(1,1;\frac{3}{2},2;-x\right)=\frac{\sqrt{\pi}}{x}\int\limits_0^{\sqrt{x}}\erfi(v)\exp\left(-v^2\right)\mathrm{d}v\quad\quad\quad(x>0),
\end{equation}
which follows directly from Eq.\ (\ref{2F2Int}). By means of some algebraic substitutions and a binomial series expansion, the latter expression can then be brought to the form

\begin{equation}
\label{2F2Expansion1}
{}_2F_2\left(1,1;\frac{3}{2},2;-x\right)=\frac{1}{2x}\int\limits_0^x\frac{1-\exp(-v)}{v\sqrt{1-\frac{v}{x}}}\,\mathrm{d}v=\frac{1}{2}\sum\limits_{m=0}^\infty\binom{-\frac{1}{2}}{m}(-1)^mx^{-(m+1)}\int\limits_0^xv^{m-1}\left[1-\exp(-v)\right]\mathrm{d}v.
\end{equation}
The last integral in the above equation for $m=0$ yields

\begin{equation}
\label{2F2Expansion2}
\int\limits_0^x\frac{1-\exp(-v)}{v}\,\mathrm{d}v=\lim\limits_{y\to0}\left[\ln(x)-\ln(y)+E_1(x)-E_1(y)\right],
\end{equation}
where $E_1(x)=\int_x^\infty\mathrm{e}^{-v}/v\,\mathrm{d}v$ is the exponential integral \cite{Abramowitz1965}. Upon taking the leading orders of the limits $y \to 0$ and $x \to \infty$ of this expression, we finally obtain

\begin{equation}
\label{2F2Expansion3}
\int\limits_0^x\frac{1-\exp(-v)}{v}\,\mathrm{d}v\sim\gamma+\ln(x),
\end{equation}
with $\gamma\approx0.577$ denoting the Euler-Mascheroni constant. For $m\geq1$ the integrand on the r.h.s of Eq.\ (\ref{2F2Expansion1}) can readily be integrated \cite{Gradshteyn2007}, namely

\begin{equation}
\label{2F2Expansion4}
\int\limits_0^x v^{m-1}\left[1-\exp(-v)\right]\mathrm{d}v=\frac{x^{m}}{m}+\exp(-x)\sum\limits_{k=0}^{m-1}\binom{m-1}{k}k!\,x^{m-1-k}\quad\quad\quad(m\ge1).
\end{equation}
In conclusion, the asymptotic expansion of the hypergeometric function of Eq.\ (\ref{2F2IntNeg}) for large negative arguments reads, to the lowest orders,

\begin{equation}
\label{2F2Expansion5}
{}_2F_2\left(1,1;\frac{3}{2},2;-x\right)\sim\frac{1}{2x}\left[\ln(x)+\gamma+\sum\limits_{m=1}^\infty\binom{-\frac{1}{2}}{m}\frac{(-1)^m}{m}\right].
\end{equation}

To sum the series of Eq.\ (\ref{2F2Expansion5}), we start from the integral representation of the digamma function $\psi(x)$ \cite{Gradshteyn2007},

\begin{equation}
\label{Digamma1}
\psi(x)=-\gamma+\int\limits_0^1\frac{v^{x-1}-1}{v-1}\,\mathrm{d}v,
\end{equation}
which, in turn, can be expanded in a binomial series, yielding

\begin{equation}
\label{Digamma2}
\psi(x)=-\gamma-\sum\limits_{m=1}^\infty\binom{x-1}{m}\frac{(-1)^m}{m}.
\end{equation}
On setting $x=1/2$ in Eq.\ (\ref{Digamma2}), one obtains the identity \citep{Abramowitz1965}

\begin{equation}
\label{Digamma3}
\sum\limits_{m=1}^\infty\binom{-\frac{1}{2}}{m}\frac{(-1)^m}{m}=-\gamma-\psi\left(\frac{1}{2}\right)=\ln(4),
\end{equation}
which, replaced into Eq.\ (\ref{2F2Expansion5}), leads to our final result,

\begin{equation}
\label{2F2Expansion6}
{}_2F_2\left(1,1;\frac{3}{2},2;-x\right)\sim\frac{\ln(4x)+\gamma}{2x}\quad\quad\quad(x\to\infty).
\end{equation}

The asymptotic expansion of the ${}_2F_2$ function for large {\it positive} arguments follows immediately from the identity

\begin{equation}
\label{2F2Expansion7}
{}_2F_2\left(1,1;\frac{3}{2},2;x\right)=\frac{\pi}{2x}\erf\left(\sqrt{x}\right)\erfi\left(\sqrt{x}\right)-{}_2F_2\left(1,1;\frac{3}{2},2;-x\right),
\end{equation}
which one derives from Eq.\ (\ref{2F2Int}) by partial integration. Hence, for $x\to\infty$,

\begin{equation}
\label{2F2Expansion8}
{}_2F_2\left(1,1;\frac{3}{2},2;x\right)\sim\frac{\sqrt{\pi}\exp(x)}{2x^{3/2}}-\frac{\ln(4x)+\gamma}{2x}.
\end{equation}
By inserting the asymptotic expansions of Eqs.\ (\ref{ErfiExpansion}), (\ref{2F2Expansion5}) and (\ref{2F2Expansion7}) into Eq.\ (\ref{SechResult}), one verifies that the singularities for $x_0\to\infty$ and $x_1\to-\infty$ cancel out as expected
and the final result simplifies to the tractable expression in Eq.\ (\ref{DeltaSech}).

\bibliographystyle{mdpi}


\bibliography{taxis_entropy}


\end{document}